\documentclass[varenna]{cimento}

\usepackage{epsfig}
\newlength{\ziffer}

\newlength{\vorzeichen}

\newcommand{\TeV}{\,\mbox{Te\kern-0.2exV}}
\newcommand{\GeV}{\,\mbox{Ge\kern-0.2exV}}
\newcommand{\mGeV}{\,\mathrm{Ge\kern-0.2exV}}
\newcommand{\MeV}{\,\mbox{Me\kern-0.2exV}}
\newcommand{\keV}{\,\mbox{ke\kern-0.2exV}}
\newcommand{\eV}{\,\mbox{e\kern-0.2exV}}

\newcommand{\ifb}{\,\mbox{fb}^{-1}}
\newcommand{\pb}{\mbox{pb}}

\newcommand{\bea}{\pagebreak[3]\begin{samepage}\begin{eqnarray}}
\newcommand{\eea}{\end{eqnarray}\end{samepage}\pagebreak[3]}
\newcommand{\beq}{\begin{equation}}
\newcommand{\eeq}{\end{equation}}

\newcommand{\abb}{Fig.~\ref}
\newcommand{\fig}{\abb}

 \newlength{\howlong}

\title{Searches for Rare and Beyond the Standard Model\\ Top~Quark Decays at the Tevatron}

\author{Daniel Wicke}

\institute{Bergische Universit{\"a}t Wuppertal -- Gau\ss{}str. 20, D-42097 Wuppertal, Germany\\
  for the CDF and D0 collaborations
}
\PACSes{%
\PACSit{14.65.Ha}{Top quarks}
\PACSit{14.70.Pw}{Other gauge bosons}
\PACSit{14.80.-j}{Other particles (including hypothetical)}
}

\begin{document}

\maketitle

\begin{abstract}
The Tevatron experiments CDF and D0 investigated the top quark decay in search
of deviations from the Standard Model. Rare processes expected in the Standard
Model as well as decays including hypothetical particles of models beyond the
Standard Model were investigated. This contribution gives an overview of the
studies that were performed with datasets of up to $7.5\ifb$ of integrated luminosity.
\end{abstract}
\section{Introduction}
Since the discovery of the top quark by CDF and D0 at the Tevatron in
1995~\cite{Abe:1995hr,Abachi:1995iq} the 
number of top events available for experimental studies has been
increased by more than two orders of magnitude. The Tevatron recently completed its
operation with a delivered integrated luminosity of more than $10\ifb$.  
Up to 75\% of these data have already been analyzed to investigate, amongst
other studies, whether the decay of top quark actually behave as expected 
by the Standard Model (SM). 
Modifications of the top quark decay may arise from modifications
affecting how the top quark acts under the electromagnetic, weak or strong
interaction. Also hypothetical particles of models beyond the
Standard Model can change the top quark decay.

To start we shall remind the reader of the results on the decay width of the
top quark. Section~\ref{sec:rare} discusses 
the investigation of rare processes due to 
electromagnetic and weak interaction. 
Analyses assuming particles beyond the SM are described in
Section~\ref{sec:bsm}.  This section finishes with a study on colour flow 
which can also be interpreted as a test of the strong interaction in top quark
decays within the SM.

\section{Width}
\label{sec:width}
The overall top quark width has been determined at the Tevatron in different
ways. CDF has investigated the distribution of reconstructed top quark masses
to obtain an upper limit on the width. Using $4.3\ifb$ of data they obtain
$\Gamma_t<7.5\GeV$~\cite{Aaltonen:2008ir,CdfNote10035}.
D0 has combined the results from the branching fraction measurment in top quark pair
production  with the measurement of the single top quark
$t$-channel production cross-section. The $t$-channel production
cross-section, measured from $2.3\ifb$,  is proportional to the partial width  $\Gamma_{t}(t\rightarrow
Wb)$. The branching fraction $R = \frac{B(t \rightarrow Wb)}{B(t \rightarrow
  Wq)}$ obtained from $1\ifb$ is then used to determine the total width of the top quark to be
$\Gamma_t=1.99^{+0.69}_{-0.55}\GeV$~\cite{Abazov:2010tm}.

These results are in good agreement with the SM expectation, but leave enough
room for modifications of rare processes and for BSM processes in the top
quark decay.
\section{Rare Processes}
\label{sec:rare}
\subsection{$tt\gamma$}
From charge conservation in its decay to SM particles, the top quark charge
could be constrained to the expected $2/3$, excluding the  exotic
$4/3$ option~\cite{Abazov:2006vd,CdfNote8967,CdfNote9939}. 
Nevertheless the top quark could have a anomalous coupling to the photon. 
\begin{figure}[b]
  \centering
\includegraphics[angle=90,width=0.5\linewidth,clip,trim=10mm 0mm 12mm 0mm]{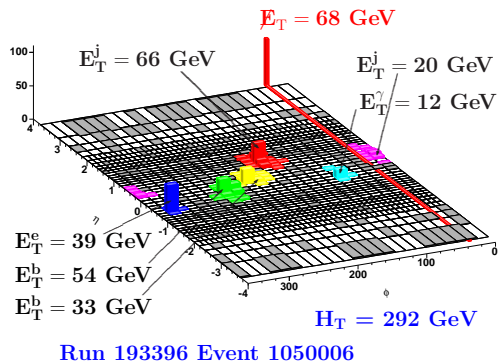}
  \caption{Top pair event with hard photon in the final state 
           observed by CDF~\cite{CdfNote10270}.}
\label{fig:ttgammaevent}
\end{figure}

CDF investigated the production of top quark pair events that include an
isolated energetic photon. Special emphasis is given to the modeling of
background due to misidentified photons, which is derived from data.
In $6.0\ifb$ of data CDF finds 30 candidate events,
c.f.~\fig{fig:ttgammaevent}. 
CDF concludes to see evidence for top pair plus photon production with
$\sigma_{t\bar t\gamma}=0.18\pm0.08\pb$~\cite{CdfNote10270}.
The result is in good agreement with the SM NLO prediction of $0.17\pb$.

\subsection{Associated Higgs Production}
In the SM the production of a Higgs boson in association with top quark pair
events is a very rare process. In models with multiple Higgs doublets or in
presence of anomalous contributions to the top quark Yukawa couplings this
process may be enhanced.

CDF and D0 have searched for this channel in semileptonic top quark decay with
various requirements on the number of jets, the number of identified $b$-jets
and with or without explicit reconstruction of the charged lepton. 
The D0 result obtained with $2.1\ifb$ is shown in \fig{fig:HiggsLimits}
(left). It is compared to predictions of a model with a heavy axigluon~\cite{d0note5739}.
CDF results are obtained using $7.5\ifb$ and $5.7\ifb$ with and without
identified leptons, respectively~\cite{CdfNote10574,CdfNote10582}. 
The obtained limits are presented in terms
of the  cross-section for $ttH$ relative to the SM expectation, \fig{fig:HiggsLimits}
(middle and right). Neither of the experiments sees an evidence for the anomalous ttH
production.

\begin{figure}[tb]
  \centering
\includegraphics[width=0.43\linewidth]{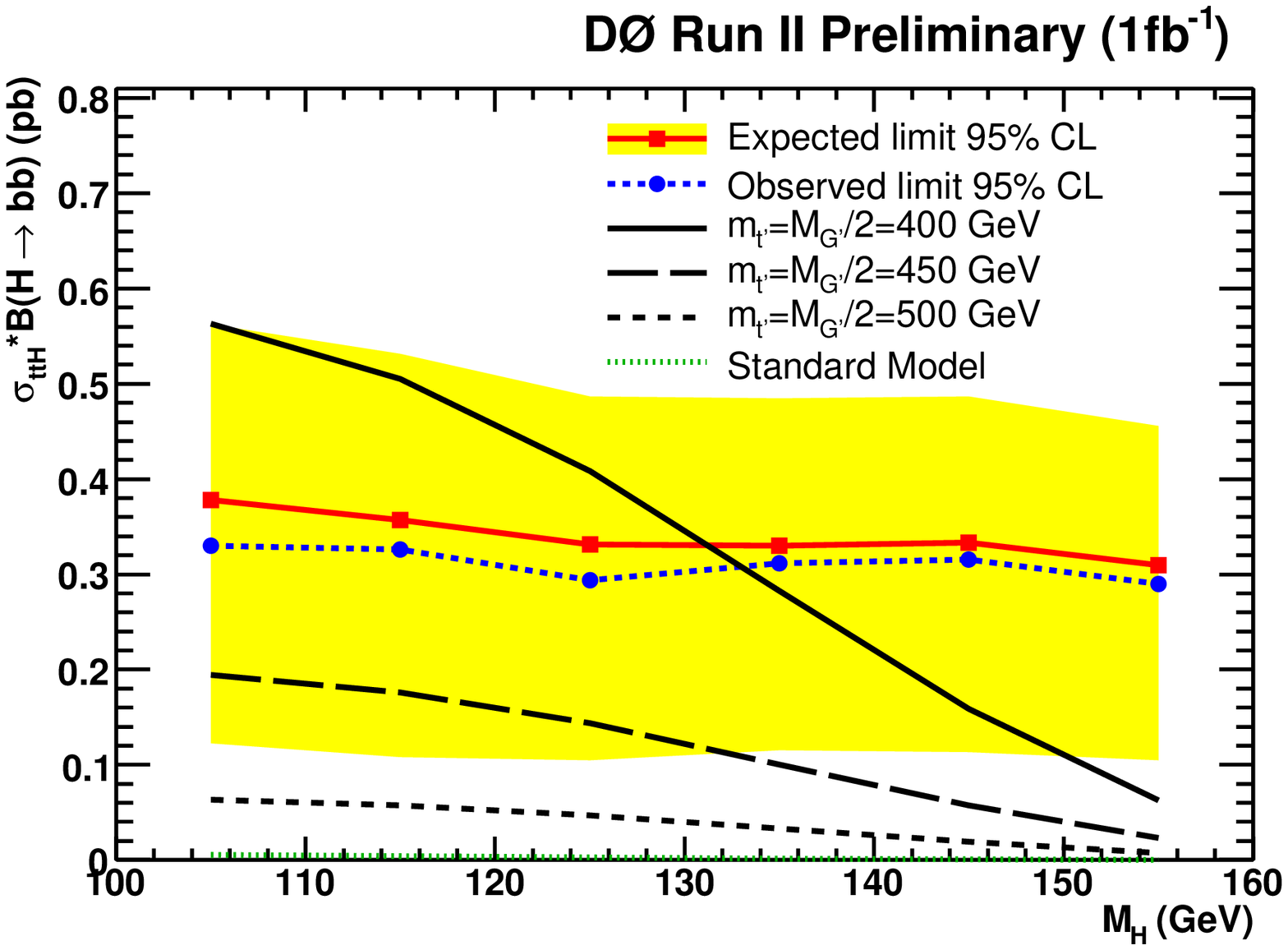}
\includegraphics[width=0.47\linewidth]{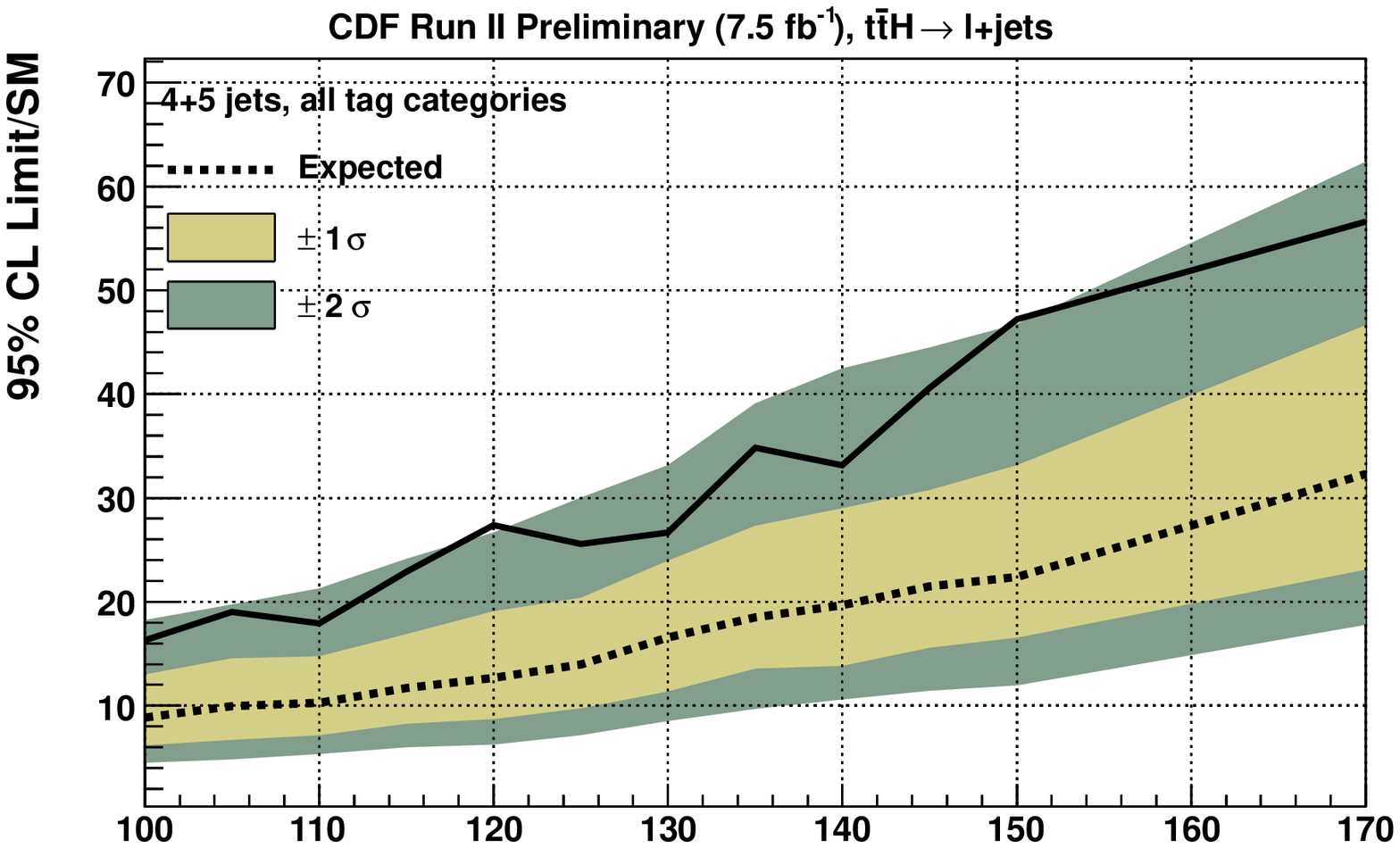}%
  \caption{%
Limits on cross-section for $ttH$ production obtained by D0~\cite{d0note5739} (left).
Limits on the ratio of the observed cross-section for $ttH$ over the
SM expectation for selection with  explicit
lepton selection determined by CDF~\cite{CdfNote10574,CdfNote10582} (right). 
}
  \label{fig:HiggsLimits}
\end{figure}

\subsection{Flavour Changing Neutral Currents}
Both Tevatron experiments also investigated their data samples for signs of top quark
decays through flavour changing neutral currents (FCNC). 
CDF investigated dilepton events assuming one FCNC and one hadronic top quark
decay  using $1.9\ifb$~\cite{Aaltonen:2008aaa}. D0 checked for trilepton
events in assuming one FCNC and one leptonic top quark decay in
$4.1\ifb$~\cite{Abazov:2011qf}. The results of both experiments are given in
terms of anomalous $tqZ$ and $tq\gamma$ couplings. These are compared to other
such results in \fig{fig:fcnc-summary}.


\begin{figure}[tb]
\null\hfill%
\begin{minipage}[t]{0.41\linewidth}
      \centering
\includegraphics[width=0.9\linewidth]{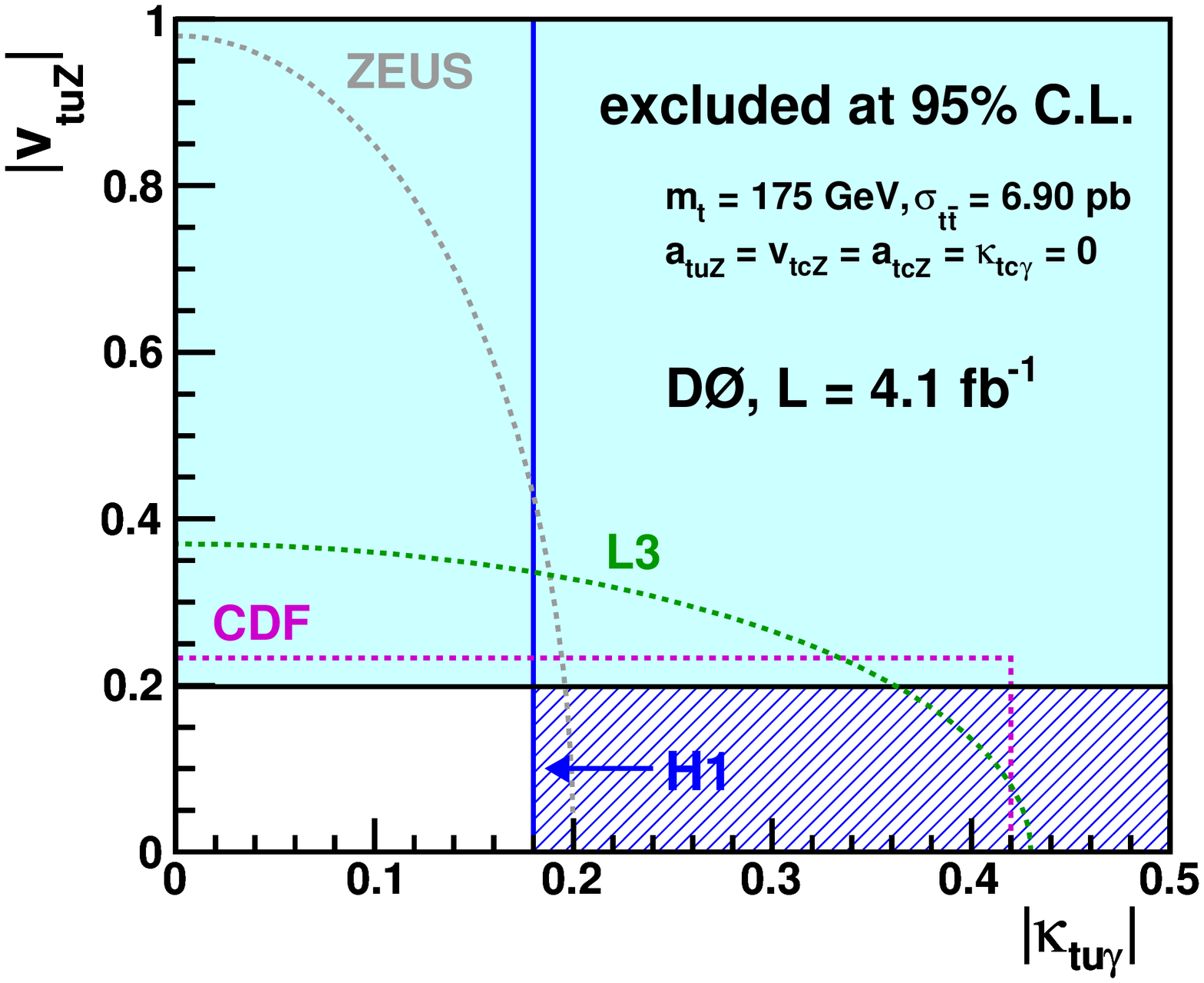}  
  \caption{Summary of limits on FCNC anomalous top quark couplings~\cite{Abazov:2011qf}.}
  \label{fig:fcnc-summary}
  \end{minipage}\hfill
  \begin{minipage}[t]{0.51\linewidth}
  \centering
\includegraphics[width=0.9\linewidth]{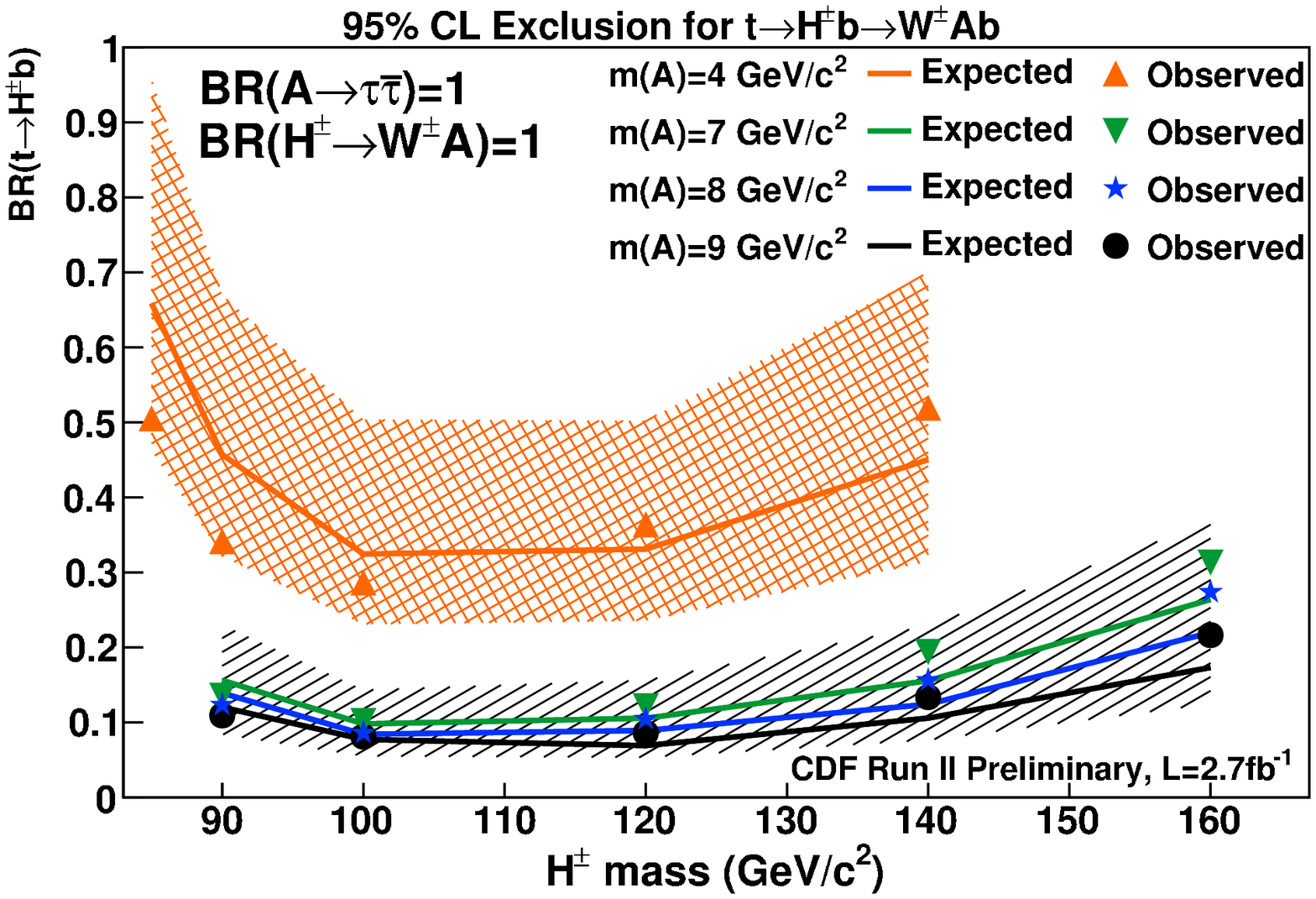}
  \caption{Limits on the contribution of a CP-odd Higgs boson to top quark decays~\cite{Aaltonen:2011aj}.}
  \label{fig:HplusOdd}
  \end{minipage}\hfill\null
\end{figure}

\section{Beyond the Standard Model Particles}
\label{sec:bsm}

\subsection{Decay to Charged Higgs Bosons}
Many BSM models predict the existence of charged Higgs bosons. 
Depending on their mass these could contribute to the top quark decay and
thereby alter the branching fractions expected from the SM, as well as the
decay kinematics.

Both Tevatron experiments have searched for charged Higgs bosons in top quark
decays. D0 provides results based on comparing the event rates obtained 
for semileptonic and dileptonic top quark pair decays including a subchannel with
one hadronic $\tau$ decay. From these the branching
fraction can be constrained to be below $15$ to $25\%$ depending on the $H^\pm$ boson mass
and its decay mode~\cite{:2009zh}.
CDF reconstructs the decay kinematic and investigates the distribution of
reconstructed  $H^\pm$ boson mass. This allows to achieve stricter limits,
if the $H^\pm$ boson mass is far from the $W$ boson mass~\cite{Aaltonen:2009ke}. 

In addition CDF has searched for a CP-odd Higgs boson in the decay chain of
$t\rightarrow H^\pm b\rightarrow W^\pm a_1^0 b$. The dominating decay channel for  $m_{a_1^0}<2m_b$ is
$a_1^0\rightarrow \tau^+\tau^-$ and yield additional soft tracks from $\tau$
decays. Using $2.7\ifb$ of data CDF sets limit on the branching fraction of
top to this decay mode for various assumed mass values of $H^\pm$ 
and $a_1^0$ as shown in \fig{fig:HplusOdd}~\cite{Aaltonen:2011aj}

\subsection{Colour Flow in Top Pair Decays}

In the SM the top quark decay happens through the colourless $W$ boson. 
Conceptually, also a coloured object  (octet) could mediate the process. 
Exploiting observables that are sensitive to the amount of hadrons produced
between two jets gives some handle to distinguish singlet from octet processes.
In semileptonic top decay from $5.3\ifb$ D0 obtains that the fraction for
production through colourless (singlet) decay to be 
$f_\mathrm{Singlet}=0.56\pm0.36_\mathrm{stat}\pm0.22_\mathrm{syst}$ consistent
with the SM expectation of 1~\cite{Abazov:2011vh}. 

\section{Summary}
The Tevatron experiments CDF and D0 have investigated  top quark decays
for various rare and BSM channel. The analyses presented use up to $7.5\ifb$
of $p\bar p$ collisions with $\sqrt{s}=1.96\TeV$. 
The results obtained probe the top quark decay down to the several percent level.
So far no significant deviation from the SM model was seen.
\bibliographystyle{varenna}
\bibliography{Dzero,CDF,local}
\end{document}